\documentstyle[aps,prl,multicol,epsf]{revtex}
\def\({\begin{equation}}
\def\){\end{equation}}
\begin{document}                
\draft
\title{Effect of Time Reversal Symmetry Breaking on the Density
of States in Small Superconducting Grains}
\author{I. S. Beloborodov$^1$, B. N. Narozhny$^{2,3}$, 
and I. L. Aleiner$^{2,3}$ }
\address{$^1$Theoretische Physik III,
Ruhr-Universit\"at Bochum, 44780 Bochum, Germany\\
$^2$Department of Physics and Astronomy, SUNY Stony Brook,
Stony Brook, NY 11794\\ 
$^3$Centre for Advanced Studies, Drammensveien 78, N-0271 Oslo, Norway} 
\date{\today}
\maketitle
\begin{abstract}
We show that in ultra-small superconducting grains any concentration of
magnetic impurities or infinitely small orbital effect of magnetic
field leads to destruction of the hard gap in the tunneling density of
states. Instead, though exponentially suppressed at low energies, the
tunneling density of states exhibits the ``soft gap'' behavior,
vanishing linearly with excitation energy, as the energy approaches
zero.
\end{abstract}


\begin{multicols}{2}

The gap in the tunneling density of states (DoS) is one of the most
fundamental manifestations of microscopic mechanism \cite{BCS} of
superconductivity \cite{gap}. The existence of this gap is closely
related to the time-reversal symmetry in the superconductor. As a result,
non-magnetic impurities (at low densities) do not affect the
gap, $\Delta$, in conventional superconductors (the Anderson theorem
\cite{and}). It is magnetic impurities that violate the symmetry of
the superconducting state and act as pair breakers. Therefore
bound states below the gap are created and the gap is suppressed 
\cite{abr}. This effect of the suppression of the gap has been
observed experimentally and it is now a textbook example \cite{tin}.

If the density of the magnetic impurities $n_m$ exceeds certain
critical value, $n_c$ the mean field transition becomes forbidden at
all.  Even at much smaller density, the magnetic impurities suppress
the DoS.  This suppression of the hard gap manifests itself in the
appearance of the finite DoS at energies (measured from the Fermi
level) $\epsilon$, smaller than the mean field  gap
$|\epsilon| < \Delta$, \cite{abr}.  However, at $n_m < 0.9 n_c$, the
self-consistent Born approximation (SCBA) of Ref.~\cite{abr} predicts
the DoS still vanishing at energies smaller than a certain value,
which is referred to as the renormalized gap $\epsilon^*$, (see
dashed line in Fig.~1.) In the region $0.9 n_c < n_m < n_c$, the
so-called regime of gapless superconductivity occurs, where the gap in
DoS vanishes even though the superfluid density still remains finite.

The prediction of SCBA about the existence of the finite gap in the
density of states even at the already broken time reversal symmetry 
is intriguing since there is no longer a symmetry reason for vanishing 
of the DoS. The situation in some sense reminds the problem of 
exponentially small tails in the DoS at small energies in normal 
materials \cite{Halperin,lif,lan}.  The appearance of such tails is
non-perturbative. It means, that any approach which consists of 
selecting some class of diagrams based on expansion in disorder 
strength will result in zero DoS under the gap. Rather, in order to 
find the shape of such tails one has to perform the instanton analysis 
(also known as the ``optimal fluctuation method'' 
\cite{Halperin,lif,lan}). In this Letter, we perform such
analysis for the effect of magnetic impurities or small magnetic field
on the DoS of superconducting grains \cite{lar}.  We will demonstrate 
that the DoS at $0 < |\epsilon |< \Delta$ becomes finite, though
exponentially small, no matter how small the magnetic field or the 
concentration of magnetic impurities is. At $|\epsilon| \to 0$, the DoS 
is shown to exhibit the ``soft gap'' behavior vanishing linearly with 
excitation energy.

The Bogolyubov equations\cite{Bogolyubov} for
quasiparticle spectrum in $s$-wave superconductors are

\begin{equation}
\epsilon \hat{\psi} = \hat{\cal H} \hat{\psi},
\label{eq:1}
\end{equation}

\noindent
where $\hat{\psi}$ is the Gor'kov-Nambu spinor\cite{Nambu}

\begin{equation}
\hat{\psi}
=\pmatrix{
u_\alpha(n)\cr
v_\alpha(n)
}_N.
\label{eq:2}
\end{equation}

\noindent
[Unless stated otherwise, Latin (Greek) indices label the 
orbital (spin) states].
The mean field Hamiltonian is

\begin{equation}
\hat{\cal H} = 
\pmatrix{
\hat{H} & \Delta \cr
\Delta & - \hat{\cal T}\hat{H}\hat{\cal T} 
}_N, \quad \hat{H}=\hat{H}^\dagger.
\label{eq:3}
\end{equation}

\noindent
where the one-particle Hamiltonian $\hat{H}$ may act both on the
orbital and spin coordinates of the electron, and $\hat{\cal T}$ is
the time inversion operator, $\hat{\cal T} u_\alpha(n) =
\sigma^y_{\alpha\beta} u_\beta^\ast(n)$, where $\sigma^y$ is the Pauli
matrix, $\hat{\cal T}^2 = 1$.

The crucial characteristic of the system is the symmetry of $\hat{H}$
with respect to the time inversion, 

\begin{equation}
\hat{H}= \hat{H}^s + \hat{H}^a; \quad
{\cal T}\hat{H}^s{\cal T}=\hat{H}^s;\quad
{\cal T}\hat{H}^a{\cal T}=-\hat{H}^a.
\label{eq:4}
\end{equation}

\noindent
If the time inversion symmetry is preserved, $\hat{H}^a=0$, both
diagonal entrees of the Hamiltonian (\ref{eq:3}) can be diagonalized
simultaneously, 

\begin{equation}
\hat{H}^s= 
\hat{\xi}={\mathrm diag}(\xi_1, \xi_2, \dots),
\label{eq:40} 
\end{equation}

\noindent
and one obtains the eigenvalues of Hamiltonian $\hat{H}^s$

\begin{equation}
\epsilon_i = \pm \sqrt{\xi_i^2 + \Delta^2},
\label{eq:5}
\end{equation}

\noindent
so that the hard gap in the one-particle excitation spectrum
exists independently of further model assumptions on
$\hat{H}^s$ \cite{and}. However, if the time inversion symmetry 
is broken, the answer is
not universal and we need to further specify the model. We adopt
$\hat{H}^s$ and $\hat{H}^a$ to be independent $M\times M, \ (M \to
\infty )$ Random Matrices \cite{RMT}

\begin{equation}
\langle \left| H_{ij}^s\right|^2\rangle=
M\left(\frac{\delta_1}{\pi}\right)^2,
\quad
\langle \left| H_{ij}^a\right|^2\rangle=
\frac{1}{\gamma}=
\frac{1}{\tau_H}\left(\frac{\delta_1}{2\pi}\right),
\label{eq:6}
\end{equation}

\noindent
satisfying the constraint (\ref{eq:4}). (We will omit the spin indices
where it does not cause any confusion.)
Here $\delta_1=\langle \xi_{i+1} - \xi_{i} \rangle \ll \Delta$ is the
mean level spacing, the parameter $\tau_H^{-1} \lesssim \Delta$ 
characterizes the strength of pair-breaking potential (see below), and
$\langle\dots\rangle$ stands for the ensemble averaging. Therefore,
$\hat{H}^s$ belongs to either orthogonal or symplectic ensembles and
$\hat{H}^a$ describes the crossover to the unitary ensemble.
In what follows, we choose the basis of the eigenstates of the
Hamiltonian $\hat{H}^s$, so that $\hat{H}^s$ has the form
(\ref{eq:40}) while $\hat{H}^a$ in this basis is a random
matrix with the correlation function (\ref{eq:6}).

The DoS in the system is expressed in terms of the disorder averaged 
Green function

\begin{equation}
\langle \nu(\epsilon)\rangle = -\frac{1}{2\pi} {\mathrm Im}
{\rm Tr}\langle G^{R}(\epsilon)\rangle, \; \;
\hat{G}^{R}(\epsilon)=\frac{1}{\epsilon - \hat{\cal H}+i0}.
\label{eq:7}
\end{equation}

\noindent
If the time reversal symmetry is preserved, $\tau_H \to \infty$,
the DoS is given by the usual BCS
expression

\[
\langle \nu(\epsilon)\rangle=\frac{1}{\delta_1} 
{\mathrm Re}\frac{|\epsilon|}{\sqrt{\epsilon^2-\Delta^2}}.
\]

Before we proceed, let us discuss the physical situations for which
the random matrix description (RMT) (\ref{eq:6}) is applicable. 
It is well known \cite{alt}
that RMT description of the spectrum requires all
the relevant energy scales to be much smaller than the Thouless energy
$E_T$. It translates to the condition that the size of the grains $R$ 
to be much smaller than the coherence length $\xi$, see 
Ref.~\cite{expt} for recent experiments on such grains.  
The parameter $\tau_H^{-1}$, can be related to the physical
characteristics of such grains. If magnetic field is applied it
penetrates through the small grain without screening  and  

\[
\tau_H^{-1} = E_T\left(\Phi / \Phi_0 \right)^2,
\]

\noindent
where $\Phi$ is the magnetic flux penetrating through the
grain and  $\Phi_0 = \hbar c / 2 e$ is the flux quantum.
In the case of the doping by magnetic impurities $\tau_H$ is 
given by the spin-spin scattering time

\[
\tau_H^{-1} = \tau_s^{-1} = \pi\nu_0 N_s |V_s|^2 S(S+1), 
\]

\noindent
where $n_s$ is the concentration of magnetic impurities, $S$ is the
impurity spin, $V_s$ is the scattering matrix element and $\nu_0$ is
the thermodynamic density of states. Finally, the possibility to 
neglect the non-Gaussian correlations of the 
Hamiltonian $\hat{H}^a$ is guarded by the requirement that no matrix
element of $\hat{H}^a$ exceeds the Thouless energy. As we shall see
below, the characteristic value of the matrix elements contributing to
the low energy tail of the DoS is of the order of $\Delta \ll E_T$, so
that the Gaussian approximation is justified. 
(All of these assumptions definitely break down in bulk systems
$R \gg \xi$.)

The exponentially small tails in the DoS were considered in
Refs.~\cite{Halperin,lif,lan}.  The idea is to look for the
fluctuations of the random potential $\hat{H}^a$ which form a low
energy bound state, thus leading to non-zero DoS at such energy. The
probability to form such a bound state is determined by the
distribution of the matrix elements $\hat{H}^a_{ij}$ and, while
exponentially small, should be maximized by choosing the ``optimal''
fluctuation of the potential. The resulting DoS is then proportional
to this probability.

To gain some intuition about the form of the optimal fluctuations, let
us consider the simplest possible realization of the random potential
$\hat{H}^a$ where it  couples only two eigenstates $i_0,\ j_0$ of the
Hamiltonian $\hat{H}^s$:

\begin{equation}
\hat{H}^a_{ij}=\left\{
\matrix{
iV; & i=i_0,\ j=j_0; \cr
-iV; & i=j_0,\ j=i_0; \cr
0; & \mbox{otherwise}. 
}
\right.
\label{eq:8}
\end{equation}

\noindent
All states with $i\neq i_0,j_0$ decouple, and the relevant Hamiltonian
reduces to  $4 \times 4$ matrix:

\begin{mathletters}
\label{eq:9}
\begin{eqnarray}
&\hat{\cal H}^{(0)} = 
\pmatrix{\hat{\cal H}_{i_0 i_0} & \hat{\cal H}_{i_0 j_0}
\cr
\hat{\cal H}_{j_0 i_0} & {\cal H}_{j_0 j_0}
 },& \nonumber\\ 
&\hat{\cal H}_{i_0 j_0} =
i V \hat{I}_N, \quad 
{\cal H}_{j_0 i_0} = \hat{\cal H}_{i_0 j_0}^*&
\label{eq:9a}
\\
&\hat{\cal H}_{i_0 i_0}=\xi_{i_0}\hat{\sigma}_z^N +
\Delta\hat{\sigma}_x^N; \quad
\hat{\cal
H}_{j_0 j_0}=\xi_{j_0}\hat{\sigma}_z^N + \Delta\hat{\sigma}_x^N.&
\label{eq:9b}
\end{eqnarray}
\end{mathletters}

\noindent
where $\hat{I}_N$ is the 
$2\times 2$ unit matrix
in Gor'kov-Nambu space, and $\hat{\sigma}_{x,z}^N$
are the Pauli matrices in this space.
The eigenvalues of the matrix (\ref{eq:9}) are

\begin{eqnarray}
&&\epsilon_{1,2} = \pm
 \left[ 
\left(\Delta -
|V| +\frac{\xi_+^2}{2\Delta}\right)^2
+\xi_-^2
\right]^{1\over{2}},
\label{eq:10}
\\
&&\epsilon_{3,4} = \pm \left[ \Delta +
|V| + {\cal O}(\xi_{\pm}^2)\right],
\quad \xi_\pm=\frac{\xi_{i_0}\pm \xi_{j_0}}{2}.
\nonumber
\end{eqnarray}

\noindent
Two of these 
eigenvalues, $\epsilon_{3,4}$, lie above the gap, and are not
interesting for our purposes. The other two, $\epsilon_{1,2}$,
correspond to the bound states under the gap. Clearly, by
suitable choice of $V$ (and $\xi$'s), we can tune $\epsilon_{1,2}$
to the desired energy $\epsilon$. The averaged DoS
is proportional to the probability $P(V)$ to find such value of the 
matrix element. According to Eqs.~(\ref{eq:6}) and (\ref{eq:8}), we
have

\[
P(V) \propto \exp\left(-\frac{\gamma}{2}V^2 \right)
\]

\noindent
The minimal value of $V$, providing the level ~(\ref{eq:10}) to
have energy $\epsilon$, is $|V| = \Delta - |\epsilon|$, and we obtain

\begin{equation}
\langle\nu(\epsilon)\rangle \propto
\exp\left[-\frac{\gamma}{2}\left(|\epsilon| - \Delta\right)^2 \right]
\label{eq:11}
\end{equation}

\noindent
where we omit all pre-exponential factors (to be calculated below). 

Equation (\ref{eq:11}) is the main physical result of 
this Letter. We have shown, that the DoS in the ultra small 
superconducting grain possesses the exponentially small tail at low
energies even below the renormalized value of the gap  obtained using
the self-consistent Born approximation. 

To make our derivation rigorous we have (i) to prove that the ansatz
(\ref{eq:8}) is indeed a saddle point in the ensemble averaging; (ii)
to calculate the pre-exponential factor by summing over all saddle
points $(i_0,j_0)$ and integrating over the fluctuations
around the saddle point. 


{\em Saddle point} --- To find the saddle point one has to minimize 
the exponent of the Gaussian probability (\ref{eq:6})

\begin{equation}
{\cal E} = \frac{\gamma}{2}\sum_{i > j}\left|H^a_{ij} \right|^2
\label{eq:12}
\end{equation}

\noindent
with respect to  all the matrix
elements $H_{ji}^a$, subjected to constraint (\ref{eq:4}) and the
condition

\begin{equation}
\epsilon = \epsilon_0\{\hat{H}^a\}
\label{eq:13}
\end{equation}

\noindent
where $\epsilon_0\{\hat{H}^a\}$ is the smallest eigenvalue of the
Hamiltonian (\ref{eq:1}). 
This involves finding a solution
of the equations

\begin{equation}
\frac{\partial}{\partial {H}^a_{ij}}
\left[{\cal E}
+ \sum_{i'< j'}{\Lambda }_{i'j'}\left({H}^a_{i'j'}+{H}^a_{j'i'} \right)
 + \lambda \epsilon_0\{\hat{H}^a\}
\right]
=0
\label{eq:14}
\end{equation}

\noindent
for all $1 \leq i < j\leq M$. Here $\Lambda_{i'j'}$ and $\lambda$ are
the Lagrange multipliers to be found from the conditions (\ref{eq:4})
and (\ref{eq:13}). Excluding $\Lambda_{ij}$, we find from 
Eq.~(\ref{eq:14})

\begin{equation}
{\bar{H}^a_{ij}} = \frac{\lambda}{2\gamma}
\left[
\frac{\partial \epsilon_0}{\partial {H}^a_{ij}}
- \frac{\partial \epsilon_0}{\partial {H}^a_{ji}}
\right].
\label{eq:15}
\end{equation}

\noindent
where

\[
\frac{\partial \epsilon_0}{\partial {H}^a_{ij}}
= \tilde{u}^\ast(i)\tilde{u}(j)+\tilde{v}(i)^\ast\tilde{v}(j).
\]

\noindent
Here $\tilde{u}(n),\ \tilde{v}(n)$ are the components of the Nambu
spinor $\hat{\psi}_j$ [see Eq.~(\ref{eq:2})], 
corresponding to the eigenstate $\epsilon_0$. 
Substituting Eqs.~(\ref{eq:15}) and (\ref{eq:40}) into 
Eqs.~(\ref{eq:1}), (\ref{eq:3}),
we obtain

\begin{eqnarray}
&&
{\; \; \; \; \; \; \; \; \; \; }
\epsilon \hat{\psi}_j
= \left(\xi_j \hat{\sigma}_z + 
 \Delta \hat{\sigma}_x + \hat{A}
\right)\hat{\psi}_j
- \hat{B}\hat{\psi}_j^*,
\nonumber\\
&&\hat{A}= \frac{\lambda}{2\gamma}\sum_n \hat{\psi}^\dagger_n \otimes \hat{\psi}_n,
\quad \hat{B}= \frac{\lambda}{2\gamma}
\sum_n \hat{\psi}_n^T \otimes \hat{\psi}_n,  
\label{eq:16}
\end{eqnarray}

\noindent
which, together with the normalization condition,

\[
\sum_j\hat{\psi}^\dagger_j\hat{\psi}_j  =1
\]

\noindent
constitute the matrix analogue of the non-linear Schr\"{o}dinger
equation of Ref.~\cite{lan}.

The essential simplicity of the random matrix model (\ref{eq:16}) stems
from the independence of the non-linear terms $\hat{A},\hat{B}$ of the
state index $j$. Namely, that Eq.~(\ref{eq:15}) can be considered as a 
linear equation for a state $j$, while the coefficients $\hat{A},\hat{B}$ 
have to be found self-consistently. With non-linear terms $\hat{A},\hat{B}$ 
fixed, the non-trivial solution to Eq.~(\ref{eq:15}) for a given eigenvalue
$\epsilon$ exists only for two values of $\xi_j$ [similar to Eq.~(\ref{eq:5})].
It means, that at most only two states can be mixed. 
According to Eq.~(\ref{eq:15}), it indicates that
ansatz (\ref{eq:9}) we adopted from the very beginning is the only
possible form of the saddle point.

{\em Pre-exponential factor} --- Having convinced ourselves, that we have
found the optimal fluctuation and, thus, the exponent (\ref{eq:11})
correctly, we turn to the calculation of the pre-exponential factor in
this expression.  Our starting point is once again the optimal fluctuation
(\ref{eq:9}), however, we wish to take into account all the other
matrix elements $H_{mj}^a$ which couple the states $m=i_0, j_0$ with
all other states ($j\neq i_0, j_0$).  Since the exponential tail comes
from the two levels $i_0, j_0$, coupled by the large matrix element
$V\gg 1/\sqrt{\gamma}$, the elements $H_{mj}^a$ may take only its
typical value $\simeq 1/\sqrt\gamma \ll \Delta$.  Therefore, all of
them can be treated in perturbation theory. To the second order in
$H_{mj}^a$ the effective Hamiltonian acting in the reduced Hilbert space
of the two states $i_0, j_0$ acquires the form (\ref{eq:9}) with the
entrees changed due to the mixture of all the other levels
(indices $m,n =i_0,j_0$) :

\begin{eqnarray}
&&\hat{\cal H}^{eff}_{mn} = \hat{\cal H}^{(0)}_{mn}
+ \sum\limits_{j\ne i_0,j_0}
H_{mj}^a 
\frac{\epsilon\hat{I}_N + \xi_j\hat{\sigma}_z^N + \Delta\hat{\sigma}_x^N }
{\epsilon^2 - \xi^2_j - \Delta^2}
H_{jn}^a \label{eq:17}
\end{eqnarray}

Substituting Eq.~(\ref{eq:17}) into Eq.~(\ref{eq:7}) we obtain the
contribution of two lowest levels into the (non-averaged) DoS:

\begin{eqnarray}
&&
\nu(\epsilon)=|\epsilon|(1-Z)
\delta\left[\epsilon^2 (1 - Z)^2 - W^2\right];
\label{eq:18}\\
&&W^2 = \left[(1+Z)\Delta + \frac{\xi_+^2}{2\Delta}
-\left|
H^a_{i_0j_0}
\right|\right]^2 
+{\xi_-^2}+Y^2,
\nonumber
\end{eqnarray}

\noindent
where $\xi_\pm$ are given by Eq.~(\ref{eq:10}),
and we introduced 

\[
Z = \sum \limits_{j\neq i_0, j_0} \frac{
\left|H^a_{i_0j}\right|^2 + \left|H^a_{j_0j}\right|^2
}
{2\left(\epsilon^2 - \xi_j^2 - \Delta^2\right)}, \; \;
Y =\sum \limits_{j\neq i_0, j_0}
\frac{\xi_j H^a_{i_0j}H^a_{j_0j}}{\epsilon^2 - \xi_j^2 - \Delta^2}.
\]

\noindent
For the optimal fluctuation (\ref{eq:8}), the
argument of the $\delta$ -function reproduces the spectrum (\ref{eq:10}).

The DoS (\ref{eq:18}) should be averaged over the fluctuations of the 
matrix elements and summed over states $i_0, j_0$

\begin{equation}
\langle\nu (\epsilon )\rangle
= \sum_{i_0,j_0} \prod_{i<j}
\left[
\int \frac{d \left({\mathrm Im}\,H^a_{ij}\right)}
{\sqrt{2\pi\gamma^{-1}}}
\exp\left(-\frac{\gamma}{2}\left|H^a_{ij}\right|^2 \right)
\right] \nu (\epsilon ).
\label{eq:19}
\end{equation}

\noindent
At energies $\epsilon\gg \delta_1$, we can neglect
the level repulsion and replace the sum over $i_0, j_0$ by the integral

\[
\sum_{i_0,j_0} \to \frac{1}{2 \delta_1^2} \int d\xi_{i_0}d\xi_{j_0},
\]

\noindent
where factor of $1/2$ excludes double counting the same configurations
$(i_0,j_0)$ and  $(j_0,i_0)$.
Then straightforward integration in Eq.~(\ref{eq:19}) utilizing the
condition $|\epsilon| \gg \delta_1/(\Delta\tau_H)$ yields the averaged
DoS:

\begin{eqnarray}
&&
\langle\nu(\epsilon)\rangle = \frac{1}{\delta_1}
F_1(\epsilon)\exp\Big[-\pi F_2(\epsilon)\Big]  
\label{eq:20}
\\ 
&&F_1(\epsilon) = \frac{2|\epsilon|}{
\sqrt{\delta_1 \left(\Delta+ 4|\epsilon|\tau_H 
\sqrt{\Delta^2-{\epsilon}^2 }\right)}}
\left(\frac{\Delta+|\epsilon|}{\Delta-|\epsilon|}\right)^{1\over{4}}; 
\nonumber \\ 
&&
F_2(\epsilon) = \tau_H\frac{\left(\Delta - |\epsilon|\right)^2}{\delta_1}
- \frac{5}{4}\sqrt{\frac{\Delta^2-\epsilon^2}{\delta_1^2}}.
\nonumber
\end{eqnarray}

Equation (\ref{eq:20}) is the main quantitative result of this Letter.
It gives the parametrically exact description of the exponential tail
in the DoS.  It is valid as long as the exponent $F_2$ is larger than
unity. At low energies $\epsilon \ll \Delta$  we can neglect
the second term in $F_2(\epsilon)$, reproducing  the qualitative result
Eq.~(\ref{eq:11}).

\begin{figure}
\epsfysize = 5.5cm
\vspace{0.2cm}
\centerline{\epsfbox{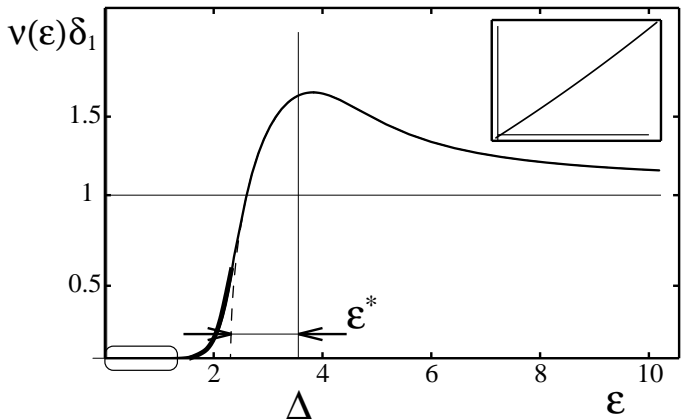}}
\vspace{0.2cm}
\caption{DoS for a superconducting grain with broken time reversal
symmetry. The thick solid line is the tail of the DoS 
~(\protect\ref{eq:20}). 
The dashed line represents the result of the self-consistent Born
approximation Ref.~\protect\onlinecite{abr}.
The inset shows the behavior of the tail at $\protect\epsilon\to 0$.
The curves are plotted for $\protect\Delta = 3.5 \delta_1$ and
$\protect\tau_H = 1.5\delta_1^{-1}$. }
\end{figure}

At larger energies, when $\epsilon$ approaches $\Delta$, the validity
of our considerations and of Eq.~(\ref{eq:20}) breaks down at the
point $\epsilon^*$ where the two terms in the exponent $F_2(\epsilon)$
become of the same order. Remarkably, that the point
$\Delta- |\epsilon^*| \simeq \Delta/(\Delta\tau_H)^{2/3}]$ is
parametrically the same as the ``renormalized gap'', $\epsilon^*$,
predicted by SCBA \cite{abr}. Combining
these results, we can describe the DoS in the grain by the continuous
function depicted on Fig.~1.

At small energies $\epsilon\to 0$ result (\ref{eq:20}) 
vanishes linearly due to the pre-exponential factor $F_1(\epsilon)$. 
The appearance of this linear suppression is easy to reveal already on
the level of qualitative analysis. Indeed, one can see from
Eq.~(\ref{eq:11}) that the levels near the $\epsilon = 0$ repel each
other due to the difference in energies of the one electron state
$\xi$. It means that the contributions of the levels is always limited
by $|\xi_i - \xi_j| < |\epsilon|$ which gives the corresponding
smallness in the integration domain. Finally, $|\epsilon| \lesssim
\delta_1$, the level repulsion \cite{RMT} between those orbits should be taken
into account which results in the additional suppression by a factor
$(\epsilon/\delta_1)^\beta$, where $\beta = 1 (4)$ in the absence
(presence) of the spin-orbit coupling.
 
To conclude, we have shown that in a small superconducting grain
breaking the time reversal symmetry leads to the appearance of the 
exponentially small tail in the DoS at energies smaller than the
BCS gap. The DoS is non zero for all energies $|\epsilon|<\Delta$
except for the point $\epsilon = 0$, where the DoS vanishes linearly
in energy. Thus the grain no longer exhibits the hard gap in the
excitation spectrum as predicted by Ref.~\onlinecite{abr}. Our 
results are non perturbative as they were obtained by considering 
the optimal fluctuation of the random potential.

We acknowledge discussions with A.I. Larkin and K.B. Efetov.  I.A. 
is a Packard research fellow. I.A. and I.B. acknowledge the support by
SFB 237 ''Unordnung und Grosse Fluktuationen''. The work of I.B. was 
supported by the Graduiertenkolleg 384.

\end{multicols}

\end{document}